\documentclass[prb,aps,nobalancelastpage,amssymb,twocolumn,citeautoscript]{revtex4}

\usepackage{graphicx}

\begin{document}
\preprint{PRB/Xu \textit{et al}.}

\title{Effect of Selenium doping on the superconductivity of Nb$_2$Pd(S$_{1-x}$Se$_x$)$_5$}

\author{C. Q. Niu$^1$, J. H. Yang$^{1,*}$\footnote[1]{Electronic address: yjhphy@163.com}, Y. K. Li$^1$, Bin Chen$^{2,1}$, N. Zhou$^1$, J. Chen$^1$, L. L. Jiang$^1$, B. Chen$^1$, X. X. Yang$^1$, Chao Cao$^1$, Jianhui Dai$^1$, and Xiaofeng Xu$^{1,*}$\footnote[2]{Electronic address: xiaofeng.xu@hznu.edu.cn}}
\affiliation{$^{1}$Department of Physics and the Quantum Matter Key Laboratory of Hangzhou, Hangzhou Normal University, Hangzhou 310036, China\\
$^{2}$Department of Physics, University of Shanghai for Science $\&$ Tehcnology , Shanghai,
China\\}

\date{\today}

\begin{abstract}
We study the isovalent substitution effect by partially introducing Se on S site in the newly
discovered superconductor Nb$_2$PdS$_5$ ($T_c\sim$6 K) whose upper critical field is found to be
far above its Pauli paramagnetic limit. In this Nb$_2$Pd(S$_{1-x}$Se$_x$)$_5$ (0$\leq$$x$$\leq$0.8)
system, superconductivity is systematically suppressed by the Se concentration and ultimately
disappears when $x\geq$ 0.5, after which a semiconducting-like ground state emerges. In spite of
the considerably reduced $T_c$ with Se doping, the ratio of the upper critical field $H_{c2}$ to
$T_c$, remains unaffected. Moreover, the size of the heat capacity jump at $T_c$ is smaller than
that expected for a BCS superconductor, implying that a strong-coupling theory cannot be the origin
of this large upper critical field. In addition, the low-lying quasiparticle excitations are
consistent with a nodeless gap opening over the Fermi surface. These results combined impose severe
constraints on any theory of exotic superconductivity in this system.
\end{abstract}

\maketitle

A superconductor with a remarkably large upper critical field relative to its $T_c$ always attracts
sustained interest from both experimental and theoretical
communities\cite{Lee06,Lei12,Yuan09,Okuda80}. As a notable example, a very high $H_{c2}$ observed
in the iron pnictide LaFeAsO$_{0.89}$F$_{0.11}$ was effectively ascribed to a two-band effect
\cite{Hunte08}. Theoretically, a magnetic field destroys superconductivity by two distinct
mechanisms: the \textit{orbital} effect and the Pauli paramagnetic pair-breaking effect
\cite{Uji01}. In the former, the vortices penetrate into the superconductor and the associated
supercurrent increases the kinetic energy of the system. When this kinetic energy gain exceeds the
condensation energy, the normal state recovers. In parallel, Cooper pairs may also be broken by
Zeeman splitting produced by the magnetic field coupling to the electronic spins in a spin-singlet
superconductor. In weak coupling BCS theory, this Pauli limiting field $H_p$ is $\sim$ 1.84
$T_c$\cite{Zuo00}. However, a growing number of superconductors are found to have $H_{c2}$
significantly higher than the Pauli limit\cite{Mizukami11,Lee00,Mercure12}. The excess of $H_{c2}$
beyond the Pauli limit has been accounted for by different theories such as strong spin-orbit
coupling, strong-coupling modes, multi-band effects or even spin-\textit{triplet} pairing.

Very recently, a new transition metal-chalcogenide based compound Nb$_2$Pd$_{0.81}$S$_5$ was
discovered to be superconducting below $T_c$ $\sim$ 6.6 K \cite{Zhang13}. This superconductivity
was found to be \textit{unconventional} in the sense that its upper critical field, for field
applied along the crystallographic $b$-axis, was reported to surpass the Pauli paramagnetic limit
by a factor of 3 as $T$$\rightarrow$ 0 K. It was also suggested that this superconductivity may be
in proximity to a magnetic instability. This very high upper critical field was tentatively
ascribed to the multi-band effect or spin-triplet pairing \cite{Zhang13,Zhang132}. Whilst both
represent the likelihood for the enhanced $H_{c2}$, the experimental data are far from conclusive.
Whether spin-orbit coupling or strong electron-boson coupling play a prominent role here, both of
which can also elevate $H_{c2}$, remains to be seen. In addition, its superconducting gap symmetry,
which may also provide valuable information on the pairing character (e.g. the possibility of
triplet pairing), has not yet been determined.

With these questions in mind, we study the effect of Se doping in this
Nb$_2$Pd(S$_{1-x}$Se$_x$)$_5$ system. With the partial substitution of Se for S, the lattice
parameters (and the unit cell volume) increase monotonically with Se content, indicating that a
negative chemical pressure is induced. Meanwhile, the superconductivity is gradually suppressed and
finally disappears around $x$=0.5, above which the ground state is semiconducting-like. While $T_c$
is rather sensitive to Se doping, the reduced upper critical field, $H_{c2}/T_c$, is seen to be
robust against impurities. Furthermore, the quasiparticle excitations are indicative of a fully
gapped superconducting order parameter with weak-coupling character. Collectively, these findings
shed important light on the nature of the observed superconductivity.

Polycrystalline samples of Nb$_2$Pd(S$_{1-x}$Se$_x$)$_5$ with nominal Se content of $x$=0, 0.1,
0.15, 0.2, 0.25, 0.3, 0.4, 0.6 and 0.8 were grown by a solid state reaction method\cite{Zhang13}.
The starting materials of  Nb(99.99\%), Pd(99.9\%), and S/Se(99.99\%) powders were mixed thoroughly
in the ratio of  2:1:7.2 in the glove box filled with Ar gas. The excess amount of S/Se is
necessary to compensate the high vapor pressure of S/Se during reaction. The pelletized mixtures
were loaded into an evacuated quartz tube which was slowly heated to 825$^{\circ}$C and kept at
this temperature for 48 hours before being quenched to room temperature. The  structure of the
polycrystalline samples were characterized by powder X-ray diffraction (XRD) at room temperature
using a Rigaku diffractometer with Cu $K_\alpha$ radiation and a graphite monochromator. Lattice
parameters were obtained by Rietveld refinements. The resistivity of each sample was measured with
a standard four-probe technique and the specific heat was measured by a long relaxation method
using a commercial Quantum Design PPMS-9 system.

\begin{figure}
\includegraphics[width=9cm,keepaspectratio=true]{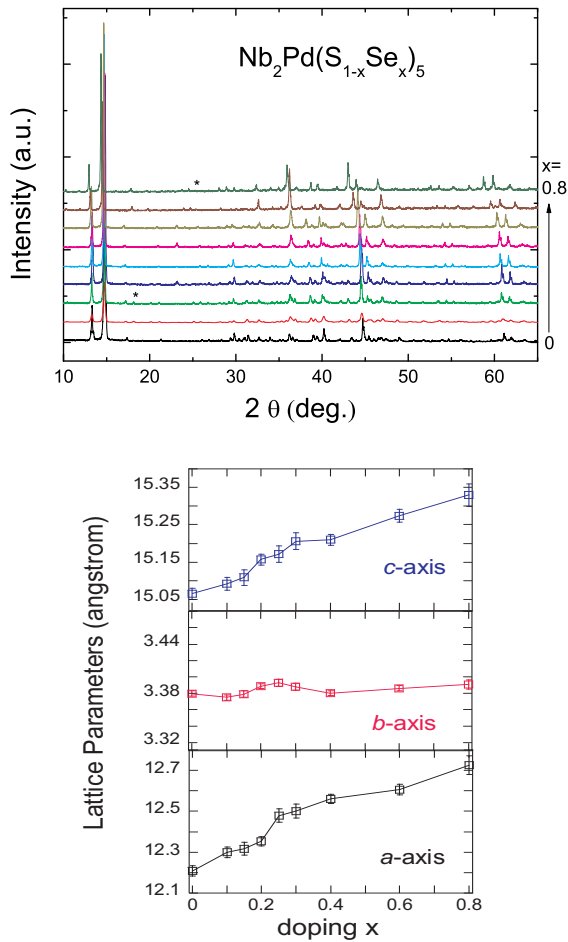}
\caption{(Color online) Top panel: The powder XRD patterns for a series of
Nb$_2$Pd(S$_{1-x}$Se$_x$)$_5$ samples studied in the paper. The asterisks mark the possible
impurity phases. Bottom Panel: The lattice parameters as a function of nominal Se content extracted
by using Rietveld analysis. } \label{Fig1}
\end{figure}

Figure 1(a) displays the powder XRD patterns of a series of Nb$_2$Pd(S$_{1-x}$Se$_x$)$_5$ samples.
The main XRD peaks of these samples can be well indexed based on a monoclinic cell structure with
the $C2/m$ space group. Extra minor peaks, marked by the asterisks in the figure, are still
detectable. Indeed, the iso-structural Nb$_2$PdSe$_5$ compound has been reported in previous
literature \cite{Keszler85}. Note that all X-ray diffractions for doped samples shift
systematically to lower 2$\theta$ angles with increasing Se concentration, implying that Se atoms
are incorporated into the lattice and lead to expansion of the lattice parameters. Figure 1(b)
shows how the extracted lattice parameters vary with Se content. Both the $c$-axis and the $a$-axis
increase monotonically with increasing Se content, while the $b$-axis shows less doping dependence.
Overall, these results indicate that the Se atoms are successfully doped into the system.

\begin{figure}
\includegraphics[width=9cm,keepaspectratio=true]{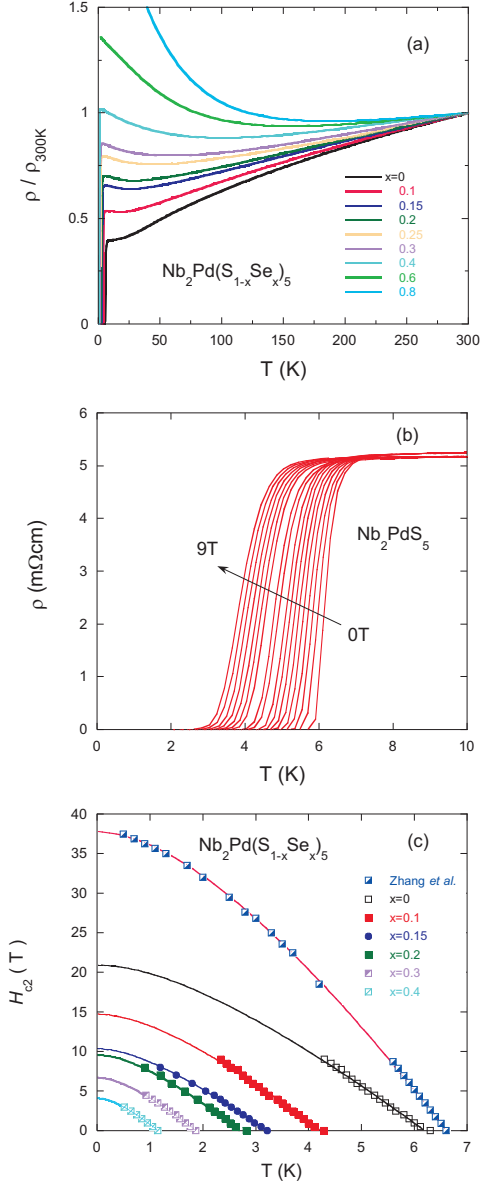}
\caption{(Color online) (a) Temperature dependence of the zero-field resistivity of
Nb$_2$Pd(S$_{1-x}$Se$_x$)$_5$ (0$\leq$$x$$\leq$0.8). All curves are renormalized to their
individual 300 K values for clarity. (b) The fixed-field temperature sweeps below $T_c$ for
Nb$_2$PdS$_5$ as an example. (c) The resultant $H_{c2}$ extracted using the criterion of 90$\%$ of
normal-state resistivity. The data by Zhang \textit{et al.} on Nb$_2$Pd$_{0.81}$S$_5$ single
crystal ($H\parallel b$) were also incorporated for comparison. The solid lines represent the
corresponding WHH fitting. } \label{Fig2}
\end{figure}

Figure 2(a) presents zero-field resistivity $\rho(T)$ curves from room temperature to the lowest
temperature studied for all samples, renormalized to their individual 300 K values for clarity. For
the parent compound Nb$_2$PdS$_5$, $\rho(T)$ is metallic on cooling from room temperature and
becomes superconducting below $T_c$ $\sim$ 6.3 K. Upon Se doping, a well-defined resistivity upturn
develops in the normal state at $T_{\rm{min}}$, which shifts to higher temperature with increasing
Se content. Meanwhile, $T_c$ is gradually suppressed by Se concentration and finally disappears as
$x\geq$0.5. Interestingly, the resistivity upturn can not be fitted to a gap-like excitation of
$\rho=\rho_0$exp($E_g/k_BT$) nor a variable-range hopping model \cite{Xu09,Ashcroftbook}. The
residual resistivity ratio is systematically reduced by Se doping. Hence, it appears more likely
that this resistivity upturn results from a disorder-induced localization effect.

The magneto-transport of the superconducting samples, as exemplified for the parent compound
Nb$_2$PdS$_5$ shown in Fig. 2(b), was studied by fixed-field temperature sweeps. The as-determined
$H_{c2}(T)$, using the criterion of 90$\%$ of normal state values, is summarized in Fig. 2(c). The
$H_{c2}$ for field aligned along the $b$-axis of Nb$_2$Pd$_{0.81}$S$_5$ single crystal is also
incorporated in the figure for comparison. As noted in the figure, whilst the $H_{c2}$ is
significantly reduced in our polycrystal compared with its single crystal profile, consistent with
its highly anisotropic Fermi surface, it still exceeds the Pauli limit by a factor of $\sim$ 2.
Similar to data for single crystals of Nb$_2$Pd$_{0.81}$S$_5$, we also satisfactorily fitted our
experimental data with the Werthamer-Helfand-Hohenberg (WHH) model \cite{WHH66,Xu13}. This is
depicted by the color-coded solid lines in Fig. 2(c).

\begin{figure}
\includegraphics[width=8cm,keepaspectratio=true]{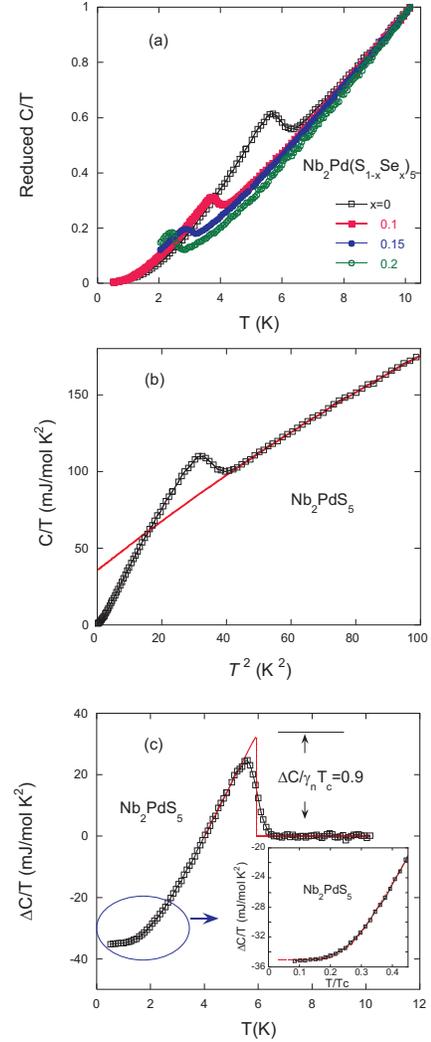}
\caption{(Color online) (a) Heat capacity anomalies associated with $T_c$ for various dopings. The
data are reduced to their 10 K values. (b) The plot of $C/T$ vs $T^2$ for Nb$_2$PdS$_5$ as a
demonstration. The red line stands for the fit to its normal-state heat capacity as $C_n/T=\gamma_n
+\beta_n T^2+\alpha_n T^4$ above $T_c$ and the extrapolation to low temperatures.
The equal entropy to the superconducting state below $T_c$ has been checked for this fitting. (c)
$\Delta C/T$ (=($C-C_n)/T$) as a function of temperature. The normalized heat capacity jump at
$T_c$, $\Delta C/\gamma_n T_c$, using the entropy conserving construction, is equal to 0.9 for
Nb$_2$PdS$_5$. The inset is the blow-up of the low-temperature data enclosed in the oval of the
main panel. The red line in the inset is the fit to $\Delta C/T$ $\sim$
$bT^{-5/2}$exp$(-\Delta_g/k_BT)-\gamma_n$. \cite{footnote1} } \label{Fig3}
\end{figure}

Fig. 3 presents the detailed calorimetric study of our samples. As seen in Fig. 3(a), a clear heat
capacity anomaly associated with the superconducting transition is observed at $\sim$ 6 K for the
parent compound. This anomaly is seen to move to lower temperature with Se doping. In Fig. 3(b),
the anomaly is isolated by subtracting the normal-state heat capacity which was fitted by
$C_n/T=\gamma_n +\beta_n T^2+\alpha_n T^4$ (the first term represents the electronic contribution,
while the remaining terms represent the phonon contribution) above $T_c$ and extrapolated to low
temperatures, as shown by the red curve \cite{Nakajima08,Nakajima12,Owen07,Walmsley13}. Although
there would be some uncertainty of the normal-state heat capacity in this method, the entropy
conservation below $T_c$ between the superconducting state and the normal state indicates that this
procedure is reliable\cite{Walmsley13}. Here, it is worth noting that the intercept of $C/T$ vs
$T^2$ as $T\rightarrow$ 0 K goes to zero, which indicates that the fraction of non-superconducting
sample is indeed very small\cite{Hussey02}. The resultant $\Delta C$, normalized to $T$, is plotted
in Fig. 3(c).

\begin{figure}
\includegraphics[width=8cm,keepaspectratio=true]{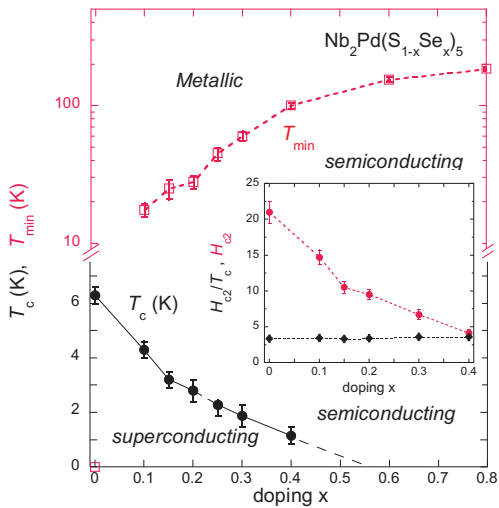}
\caption{(Color online) The phase diagram of $T_c$ and resistivity minimum $T_{\rm{min}}$ as a
function of doping level $x$. Note that the $y$-axis is broken up for $T_c$ and $T_{\rm{min}}$,
respectively. The inset shows the evolution of $H_{c2}$ and $H_{c2}/T_c$ as doping.} \label{Fig4}
\end{figure}

From Fig. 3(c), the size of the heat capacity jump in our polycrystal samples is much higher than
that reported in single crystal samples \cite{Zhang13}. Importantly, the value of the normalized
jump, $\Delta C/\gamma_n T_c$ $\sim$ 0.9 is considerably smaller than the weak-coupling BCS value
of 1.43. The value of this normalized jump was seen to vary slightly with different doping level
$x$, being the largest at a value of $\sim$ 1.1. The small $\Delta C/\gamma_n T_c$ would seem to
rule out a strong-coupling origin of the observed superconductivity\cite{Carbotte90,Popovich10},
which was reported to be responsible for the significantly enhanced $H_{c2}$ in the heavy fermion
superconductor CeCoIn$_5$\cite{Mizukami11}.

In the low temperature limit, we uncovered another interesting feature of the observed
superconductivity. In a superconductor with a gap node, one expects $C_{el}$ $\sim T^2$ so $\Delta
C/T \sim aT-\gamma_n$. Instead, a nodeless superconductor predicts $\Delta C/T$ $\sim$
$bT^{-5/2}$exp$(-\Delta_g/k_BT)-\gamma_n$\cite{Hussey02,Owen07}. As clearly seen in the inset of
Fig. 3(c), the electronic specific heat is dominated by the exponential dependence up to $T/T_c$
$\sim$ 0.45, indicating a nodeless gap on the Fermi surface. This seemingly precludes a $d$-wave
gap or a simple $p_x$($p_y$) symmetry where node(s) are present, although complex order parameter
structures such as $p_x+ip_y$ are still possible.

Finally, the resultant phase diagram is summarized in Fig. 4. Accompanied by an increase of
resistivity minimum $T_{\rm{min}}$, the superconducting transition $T_c$ is systematically
suppressed by Se doping in Nb$_2$Pd(S$_{1-x}$Se$_x$)$_5$, and ultimately disappears when $x\geq$
0.5, where the ground state is semiconducting-like, presumably induced by disorder effect.
Intriguingly, as seen in the inset, while the upper critical field $H_{c2}$ is significantly
reduced by the doping, the normalized $H_{c2}/T_c$ is rather robust against impurities, exceeding
the Pauli limiting value of 1.84$T_c$ by a factor of 2. The robustness of this large $H_{c2}$
relative to $T_c$ up to $x$ $\sim$ 0.4 seems at odds with the spin triplet origin of the enormous
$H_{c2}$ observed in Nb$_2$Pd$_{0.81}$S$_5$, which should be sensitive to non-magnetic impurity
scattering. To firmly rule out the spin triplet scenario, it would be a litmus test to study the
Knight shift suppression below $T_c$\cite{Raghu10,Lee01}.

In summary, we study the isovalent doping effect,  i.e., the partial substitution of Se for S, in
the newly discovered superconducting Nb$_2$PdS$_5$ system. While $T_c$ substantially decreases with
Se doping, its high $H_{c2}$ with respect to $T_c$, $H_{c2}/T_c$, is found to be immune to the Se
impurities. Moreover, the heat capacity study reveals that the superconductivity is fully gapped,
with relatively weak-coupling strength. Both of these findings seemingly argue against
strong-coupling and spin-triplet pairing as the origin of the large $H_{c2}$. Alternatively,
multi-band effects or strong spin-orbit coupling or even a subtle combination of the two, could be
responsible for the large $H_{c2}$. Regarding the latter, it shall be very instructive to
substitute heavier elements, such as Pt, for Pd to study the role of spin-orbit coupling in this
exotic superconductor.

The authors would like to thank N. E. Hussey, A. F. Bangura, C. M. J. Andrew, E. A. Yelland, Wenhe
Jiao, Guanghan Cao, Zhuan Xu, Xiaofeng Jin for stimulating discussions and Q. L. Ye, H. D. Wang for
collaborative support. The work is sponsored by the National Natural Science Foundation of China.

\bibliography{NbPdS}

\begin{thebibliography}{27}

\bibitem{Lee06} P. A. Lee, N. Nagaosa, X. G. Wen,  Rev. Mod. Phys. {\bf 78}, 17 (2006).
\bibitem{Lei12} H. Lei \textit{et al.}, Phys. Rev. B \textbf{85}, 094515 (2012).
\bibitem{Yuan09} H. Q. Yuan \textit{et al.}, Nature \textbf{457}, 565–568 (2009).
\bibitem{Okuda80} K. Okuda, M. Kitagawa, T. Sakakibara, and M. Date, J. Phys. Soc. Jpn. \textbf{48}, 2157–2158 (1980).
\bibitem{Hunte08} F. Hunte, J. Jaroszynski, A. Gurevich, D. C. Larbalestier, R. Jin, A. S. Sefat, M. A. McGuire, B. C. Sales, D. K. Christen, D. Mandrus, Nature {\bf 453}, 903 (2008).
\bibitem{Uji01} S. Uji, H. Shinagawa, T. Terashima, T. Yakabe, Y. Terai, M. Tokumoto, A. Kobayashi, H. Tanaka, and H. Kobayashi, Nature \textbf{410}, 908 (2001).
\bibitem{Zuo00} F. Zuo, J. S. Brooks, R. H. Mckenzie, J. A. Schlueter, and J. M. Williams, Phys. Rev. B \textbf{61}, 750 (2000).
\bibitem{Mizukami11} Y. Mizukami, H. Shishido, T. Shibauchi, M. Shimozawa, S. Yasumoto, D. Watanabe, M. Yamashita, H. Ikeda, T. Terashima, H. Kontani and Y. Matsuda, Nat. Phys. \textbf{7}, 849 (2011).
\bibitem{Lee00} I. J. Lee, P. M. Chaikin, and M. J. Naughton, Phys. Rev. B \textbf{62}, 14669 (2000).
\bibitem{Mercure12} J.-F. Mercure, A. F. Bangura, Xiaofeng Xu, N. Wakeham, A. Carrington, P. Walmsley, M. Greenblatt, and N. E. Hussey, Phys. Rev. Lett. \textbf{108}, 187003 (2012)
\bibitem{Zhang13} Q. Zhang, G. Li, D. Rhodes, A. Kiswandhi, T. Besara, B. Zeng, J. Sun, T. Siegrist, M. D. Johannes and L. Balicas, \textit{Sci. Rep.} \textbf{3}, 1446 (2013).
\bibitem{Zhang132} Q. Zhang, D. Rhodes, B. Zeng, T. Besara, T. Siegrist, M. D. Johannes, L. Balicas, arXiv:1306.6868.
\bibitem{Keszler85} D. A. Keszler, J. A. Ibers, Maoyu Shang and Jiaxi Lu, J. Solid State Chem. \textbf{57}, 68 (1985).
\bibitem{Xu09} Xiaofeng Xu, A. F. Bangura, J. G. Analytis, J. D. Fletcher, M. M. J. French, N. Shannon, J. He, S. Zhang, D. Mandrus, R. Jin, and N. E. Hussey, Phys. Rev. Lett. \textbf{102}, 206602 (2009)
\bibitem{Ashcroftbook} Ashcroft and Mermin, Solid State Physics (Cornell University Press, Cornell, 1975).
\bibitem{WHH66} N. R. Werthamer, E. Helfand, P. C. Hohenberg Phys. Rev. {\bf 147}, 295 (1966).
\bibitem{Xu13} Xiaofeng Xu \textit{et al.}, Phys. Rev. B \textbf{87}, 224507 (2013).
\bibitem{Nakajima08} Y. Nakajima, T. Nakagawa, T. Tamegai, and H. Harima, Phys. Rev. Lett. {\bf 100}, 157001 (2008).
\bibitem{Nakajima12} Y. Nakajima, H. Hidaka, T. Nakagawa, T. Tamegai, T. Nishizaki, T. Sasaki, and N. Kobayashi, Phys. Rev. B {\bf 85}, 174524 (2012).
\bibitem{Owen07} O. J. Taylor, A. Carrington, J. A. Schlueter, Phys. Rev. Lett. {\bf 99}, 057001 (2007).
\bibitem{Walmsley13} P. Walmsley \textit{et al.} Phys. Rev. Lett. 110, 257002 (2013).
\bibitem{Hussey02} N. E. Hussey, Adv. Phys. {\bf 51}, 1685 (2002).
\bibitem{footnote1} Note that the free parameters in this fitting are $b$ and $\Delta_g$ only.
$\gamma_n$ is fixed to be the one extracted from Fig. 3(b). The resulting $\Delta_g$ value is
$\sim$ 1.9$k_BT_c$, close to the weak-coupling BCS value of 1.76$k_BT_c$, further indicating that
the coupling strength is not strong in this system.
\bibitem{Carbotte90} J. P. Carbotte, Rev. Mod. Phys. {\bf 62}, 1027 (1990).
\bibitem{Popovich10} P. Popovich, A. V. Boris, O. V. Dolgov, A. A. Golubov, D. L. Sun, C. T. Lin, R. K. Kremer,and B. Keimer, Phys. Rev. lett. {\bf 105}, 027003 (2010).
\bibitem{Raghu10} S. Raghu, A. Kapitulnik, and S. A. Kivelson, Phys. Rev. Lett. \textbf{105}, 136401 (2010) .
\bibitem{Lee01} I. J. Lee, S. E. Brown, W. G. Clark, M. J. Strouse, M. J. Naughton, W. Kang, and P. M. Chaikin, Phys. Rev. Lett. \textbf{88}, 017004 (2001) .
\end{thebibliography}


\end{document}